\documentstyle[twocolumn,aps,psfig]{revtex}
\begin{document}
\draft
\flushbottom
\twocolumn[
\hsize\textwidth\columnwidth\hsize\csname @twocolumnfalse\endcsname

\title{ Electric field induced strong localization of electrons on solid
hydrogen surface: possible applications to quantum computing}
\author{Igor I. Smolyaninov }
\address{ Electrical and Computer Engineering Department \\
University of Maryland, College Park,\\
MD 20742}
\author{Vitaliy V. Zavyalov }
\address{ Kapitza Institute for Physical Problems \\
Russian Academy of Sciences \\
Moscow, Russia }

\date{\today}
\maketitle
\tightenlines
\widetext
\advance\leftskip by 57pt
\advance\rightskip by 57pt

\begin{abstract}
Two-dimensional electron system on the liquid helium surface is one of the 
leading candidates for constructing large analog quantum computers (P.M. 
Platzman and M.I. Dykman, Science 284, 1967 (1999)). Similar electron systems on 
the surfaces of solid hydrogen or solid neon may have some important advantages 
with respect to electrons on liquid helium in quantum computing applications, 
such as larger state separation $\Delta E$, absence of propagating capillary 
waves (or ripplons), smaller vapor pressure, etc. As a result, it may operate at 
higher temperatures. Surface roughness is the main hurdle to overcome in 
building a realistic quantum computer using these states. Electric field induced 
strong localization of surface electrons is shown to be a convenient tool to 
characterize surface roughness.

%\vspace{0.5cm}

%PACS number(s): 73.20.Fz; 73.20.Dx; 03.67.Lx 
\end{abstract}

]
\narrowtext

\tightenlines

Quantum computers may provide an enormous gain in the computation rate due to 
high parallelism of multi-qubit quantum evolution \cite{1}. They are viewed as a 
system of N two-level interacting quantum systems (qubits) that evolve in time 
under the action of some time-dependent Hamiltonian (computer program). The 
computer operation is based on a controlled series of qubit couplings together 
with one-qubit rotations. 

Two-dimensional electron system on the liquid helium surface is one of the 
leading candidates for constructing large analog quantum computers \cite{2}. It 
consists of a set of electrons $N \sim 10^9$ trapped in vacuum in the image 
states on the surface of a liquid helium film. The ground and the first excited 
state of an electron in the image potential may represent 0 and 1 state of a 
qubit. The electron states (qubits) and electron-electron interactions can be 
easily manipulated by external electric fields and resonant microwave radiation. 
On the other hand, these electrons are sufficiently isolated from the outside 
world: they are coupled to it only through the irregularities of the helium 
film, such as capillary waves and the substrate defects, and through the 
collisions with vapor molecules. Thus, these two-dimensional electron systems 
can behave as quantum computers with many qubits. Unfortunately, the operation 
temperature of such a computer must be below $10^{-2} K$ \cite{2} which is 
caused by small quantum state separation $\Delta E\sim 8 K$, thermal excitation 
of propagating capillary waves (or ripplons) on the liquid He surface, large He 
vapor pressure at higher temperatures, etc. 

In this Letter we are going to show that similar two-dimensional electron systems 
on the surfaces of solid hydrogen or neon may have some important advantages 
with respect to electrons on liquid helium in quantum computing applications.
Since surface roughness is the main hurdle to overcome in building a realistic 
quantum computer using these states, it is very important to develop reliable 
tools for its characterization. Electric field induced strong localization of 
electrons on solid hydrogen surface observed in the experiment is suggested as 
such a reliable tool to characterize roughness of solid hydrogen or solid neon 
surface.
 
Most of the experimental and theoretical work on two-dimensional electron
states above the dielectric surfaces with $(\epsilon -1) << 1$ has been done for 
the case of liquid helium \cite{3}. In the simplest model, the interaction 
potential $\phi $ for an electron near the surface of such a dielectric depends 
only on the electrostatic image force, and on the external electric field $E$, 
which is normal to the surface and is necessary for the electron confinement 
near the surface:

\begin{equation} 
\phi (z)=-e^2(\epsilon -1)/(4z(\epsilon +1))+eEz=-Qe^2/z+eEz 
\end{equation}

for $z>0$, and $\phi (z)=V_0$ for $z<0$, where the z axis is normal to the 
surface, and $V_0$ is the surface potential barrier. If $V_0\rightarrow \infty$ 
one obtains the electron energy spectrum:

\begin{equation} 
E_n=-Q^2me^4/(2\hbar ^2n^2 )+eE<z_n>+p^2/(2m)
\end{equation}

The first term in this expression gives the exact solution for $E=0$. The second 
term is the first-order correction for a non-zero confining field, where the 
average distance of electrons from the surface in the $n$th energy level is 
$<z_n>=3n^2\hbar^2/(2me^2Q)$. This correction provides an extremely convenient 
way of fine-tuning the energy spectrum with electric signals. In \cite{2} it was 
proposed to use patterned electrodes to adjust separately the field $E$ acting 
on each electron (qubit), so that one can address each qubit with a correctly 
chosen frequency of a microwave driving field. The maximum density of electrons 
on the dielectric surface is determined by the average external confining field: 
$n\leq E/(4\pi e)$ \cite{3}. Density values of the order of  
$10^9$ cm$^{-2}$ are routinely observed which correspond to a reasonable (from 
the point of view of current state of lithographical techniques) $\sim 300$ nm 
spacing between individual electrons (and, hence, individual electrodes 
necessary to address each of the electrons separately). The last term in (2) 
corresponds to free electron's motion parallel to the surface. 

\begin{figure}[tbp]
\centerline{
\psfig{figure=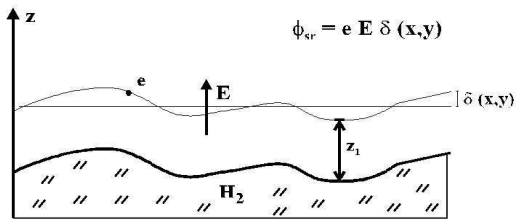,width=9.0cm,height=6.0cm,clip=}
}
\caption{ Surface topography induces variations in the random potential relief 
($H_{sr}=eE\delta (x,y)$) seeing by the electron in its lateral motion under the 
applied external confining field $E$. 
}
\label{fig1}
\end{figure}

Similar two-dimensional electron layers have been observed on the surface of 
liquid and solid hydrogen and neon \cite{3}. Resonant absorption of light for 
$1\rightarrow 2$ and $1 \rightarrow 3$ transitions in the spectrum of electrons 
levitating above the surfaces of solid hydrogen and neon has been reported and 
the frequencies of these transitions have been measured as a function of the 
confining electric field \cite{4,5}. Because of approximately three times larger 
values of $Q$ for solid hydrogen and neon, the electrons levitate closer (at a 
distance of $\sim 25\AA $) to the surface and have an order of magnitude larger 
separation $\Delta E$ between quantum energy states. Nevertheless, like 
electrons on liquid helium surface, they have pretty high mobility $\mu \sim 
10^4 cm^2/Vs$ \cite{6} limited by surface roughness. Coupling to surface 
roughness ($H_{sr}=eE\delta $, where $\delta $ represents the surface height 
variations) is supposed to determine the relaxation time $T$ from the n=2 to n=1 
state for a moving electron. It is essential to keep $T$ as large as possible 
for a successful operation of a quantum computer. Although the photoresonance 
linewidth ($\Delta \nu \sim 3\times 10^{10}$ Hz) measured in \cite{4,5} is not 
sufficiently narrow for quantum computer operation, it was most probably caused 
by inhomogeneous broadening due to variations in height of the samples and 
crystal orientation. Simple estimate of a homogeneous line width for electrons 
on atomically flat crystal surface gives six orders of magnitude smaller 
linewidth of 
$\Delta \nu \sim \omega_{12}^3d_{12}^2/(3\hbar c^3)\sim 10^4$ Hz, where $\omega 
_{12}$ and $d_{12}$ are the frequency and the average dipole moment of the 
$1\rightarrow 2$ transition. This value is just slightly larger than the value 
expected for electrons on liquid helium surface. Thus, ability to grow high 
quality atomically flat surfaces of solid hydrogen or solid neon would be of 
critical importance for building a quantum computer using surface electron 
states. Most of the other properties of electron systems on solid hydrogen and 
neon give them substantial advantages with respect to electrons on liquid helium 
in quantum computing applications. Larger separation $\Delta E$ between the 
energy states, much smaller vapor pressure at low temperatures, and the absence 
of propagating capillary waves (or ripplons) may allow a quantum computer 
operation at sufficiently higher temperatures around 4K. This would make a 
quantum computer based on surface electron states much more feasible from the 
technological point of view.       

Surface electrons have been previously suggested as a tool to characterize 
surface roughness of solid hydrogen films prepared by quench condensation on a 
glass substrate at 1.5K \cite{7}. Surface electron conductivity has been studied 
as a function of temperature, and thermal-activation-type dependencies have been 
observed for the "very rough" samples obtained. The measured activation energy 
of the order of 10K or more has been suggested as an indicator of surface 
quality of the hydrogen samples used. This method may not be appropriate for 
higher quality samples since surface roughness may itself be a function of 
temperature in the state of thermal equilibrium. The idea of alternative 
isothermic surface roughness measurements based on Anderson localization is 
illustrated in Fig.1. Variation of external confining field $E$ applied to an 
electron localized in the ground state over a rough hydrogen surface leads to 
linear variations in the random potential relief ($H_{sr}=eE\delta (x,y)$) 
seeing by the electron in its lateral motion. For a lateral roughness scale 
smaller than the electron wavelength 
($\lambda = 2\pi \hbar /(2mkT)^{1/2}\sim 40$ nm at T=10K) we may expect a 
classic case of Anderson localization \cite{8} at a sufficiently high confining 
field strength.       

According to the measurements of electron mobility on high quality crystals of 
solid hydrogen \cite{6}, at 13.4K the main contributions to surface electron 
scattering come from the collisions with hydrogen molecules in the vapor phase, 
and from the collisions with surface defects. The value of the electron mobility 
$\mu = 7500 cm^2/Vs$ measured at 13.4K is above the "minimal metallic mobility"
value $\mu _{min}$ that corresponds to the electron free propagation length $l$ 
of the order of the electron wavelength $\lambda $: the electron mobility may be 
written as

\begin{equation} 
\mu = e\tau /m = \frac{\pi e\hbar}{mkT} \frac{l}{\lambda }
\end{equation}
   
where $\tau $ is the free propagation time. At $l/\lambda \sim 1$ and T=13.4K 
the minimal metallic mobility value is equal to $\mu _{min} \approx 3100 
cm^2/Vs$. These numbers show that although the surface electrons are supposed to 
have the diffusive type of conductivity, they are not far from strong 
localization. Increase in scattering by the surface defects due to application 
of higher confining electric field should cause such a localization.     

\begin{figure}[tbp]
\centerline{
\psfig{figure=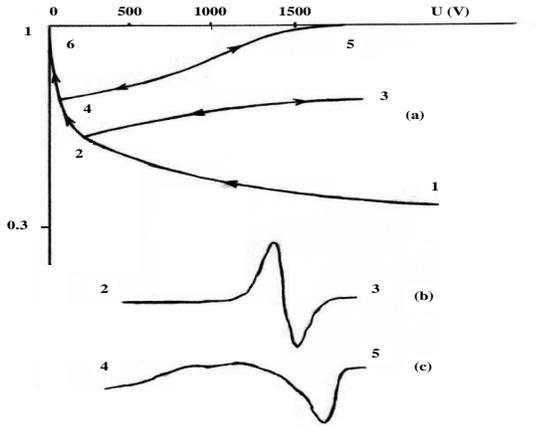,width=9.0cm,height=9.0cm,clip=}
}
\caption{(a) Radio frequency absorption of the electron layer on top of a 1 mm 
thick hydrogen crystal measured as a function of the bottom electrode potential. 
The data obtained on the way from the point (1) to points (2), (4), and (6) 
correspond to irreversible escape of the electrons from the surface with the 
electron density determined by $n = E/(4\pi e)$. The drop in absorption upon the 
confining field increase on the way from (2) to (3), or from (4) to (5) was 
reversible with the surface electron density unchanged over the duration of the 
measurements. (b,c) Photoresonance $1\rightarrow 2$ transitions detected in the electron layer on the way (2-3) (b) and (4-5) (c). Optical signal is proportional to the derivative of the optical absorption at 
$\lambda = 84.3 \mu m$. 
}
\label{fig2}
\end{figure}

An exponential drop in surface electron conductivity under the applied confining 
electric field has been indeed observed in the experiment. The method of 
conductivity measurements has been discussed in detail in \cite{4}.  
A 1.0 mm thick solid hydrogen crystal was grown on the polished sapphire 
substrate from the liquid hydrogen phase at the triple point. A two-dimensional 
electron layer levitating above the hydrogen crystal surface was created at a 
temperature of 13.4K by a pulse of electric current through a tungsten filament 
located above the crystal. Two electrodes separated by 2.3 mm produced a uniform 
confining electric field perpendicular to the crystal surface. The upper 
electrode was grounded. The lower (aluminum) electrode was deposited on the 
sapphire by vacuum sputtering. The electrodes were 30 mm in diameter and were 
large enough so that the fringe fields could be neglected near the center of the 
hydrogen surface. The lower electrode was separated by a narrow gap into two 
parts that were connected in parallel to a superconducting LC circuit with a 
resonant frequency of 94 kHz. Appearance of an electron layer on the hydrogen 
surface led to a decrease of the LC circuit's Q-factor (initially equaled to $Q 
\approx 1000$), and to the decrease of the output signal proportional to the 
non-resonant radio-frequency absorption of the electron layer. The absorption of 
the electron layer (proportional to its conductivity which is determined by the 
electrons density and the mobility of individual electrons in the layer) was 
recorded while scanning the potential of the lower electrode (see Fig.2). 
Electrons completely charged the hydrogen surface at point (1) when the pulse of 
current through the filament occurred. The data obtained on the way from the 
point (1) to points (2), (4), and (6) correspond to irreversible escape of the 
electrons from the surface with the electron density determined by $n = E/(4\pi 
e)$. The drop in absorption upon the confining field increase on the way from 
(2) to (3), or from (4) to (5) was 
reversible with the average surface electron density unchanged over the duration of the measurements. This drop in absorption is mainly caused by the increase in
the electron scattering by surface defects due to larger scattering cross-sections of the topographical defects at larger confining electric fields. 
In principle, some electron redistribution over the crystal surface may also happen, although this effect is supposed to be small in the geometry of the experiment. The surface area covered by electrons may only change by a few percents due to the fact that the horizontal component of the confining field exists only near the edges of the closely spaced top and bottom electrodes (and in the narrow gap between the electrodes and the walls of the experimental chamber). Moreover, such a redistribution of electrons would cause an opposite effect on the absorption signal due to an increase in the electron density.  

Thus, a signal proportional to the electron mobility was measured 
as a function of the confining electric field.
While the behavior of electron absorption on the way (4-5) clearly 
exhibits evidences of strong localization, the measurements (2-3) performed at 
higher electron density do not show any sign of localization within the 
measurements range. This behavior may be attributed to the two-dimensional 
screening of the lateral potential relief by the conductive electron layer. 
According to \cite{9}, the screening parameter of a non-degenerate 2D electron 
gas equals to
$S=2\pi e^2n/(kT)$, which gives the values of $S_{45}^{-1}=36$ nm and $S_{23}^{-
1}=8$ nm for the electron densities of $n_{45}=3.6\times 10^8 cm^{-2}$ and 
$ n_{23}=1.6\times 10^9 cm^{-2}$ in the measurements (4-5) and (2-3) respectively. It seems reasonable 
to assume that the screening would strongly affect the character of electron's 
conductivity when the screening length becomes comparable to the characteristic 
size of surface defects. Thus, the numbers obtained for the screening length 
indicate the characteristic lateral size (in the 8 - 36 nm range) of the surface 
roughness seen by the electrons. The fact that this lateral size is smaller than 
the electron's wavelength justifies the applicability of the Anderson 
localization model to our experimental situation. 

The data of the transport measurements were also supported by the simultaneous measurements of photoresonance $1\rightarrow 2$ transition in the electron spectrum (2). These data are presented in Fig.2(b,c). 
The optical signal proportional to the derivative of the optical absorption at $\lambda = 84.3 \mu m$ was measured using 50V modulation of the bottom electrode potential. A substantial asymmetric photoresonance line broadening can be seen as a direct result of Anderson localization. This is a natural result since each localized electron is supposed to see different local topographical environment.  
       
\begin{figure}[tbp]
\centerline{
\psfig{figure=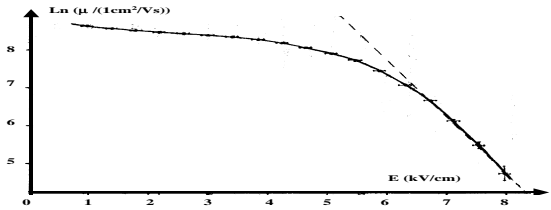,width=9.0cm,height=6.0cm,clip=}
}
\caption{ The data of the measurements (4-5) from Fig.2 plotted as a logarithm 
of electrons mobility versus confining electric field $E$. These data clearly 
show the transition from the diffusive to activation-type of conductivity of the 
surface electron layer.
}
\label{fig3}
\end{figure}

Fig.3 represents the data obtained in the measurements (4-5) plotted as a 
logarithm of electrons mobility versus confining electric field $E$. These data 
clearly show the transition from the diffusive to the activation-type of 
conductivity of the surface electron layer ($\sigma \sim exp(-E_c/kT)$), with 
the conductivity threshold $E_c$ that depends linearly on the confining field. 
This linear dependence is exactly the behavior one would expect from the model 
shown in Fig.1. Thus, the average height variation of the hydrogen surface may 
be recovered as $\delta = \frac{kT}{e}\frac{d(ln\mu )}{dE}\sim 16$ nm. The 
developed technique may be used in the growth and characterization of 
higher quality crystals of hydrogen and neon that is necessary in quantum 
computing applications. The confining field strength necessary to induce strong 
localization may serve as a convenient indicator of surface quality. 

The observed electric field induced Anderson 
localization may also be treated as a simple example of qubit coupling and de-
coupling using external electric signals, which is also very important from the 
point of view of building a large quantum computer. The degree of overlap of the 
electrons wave functions may be conveniently adjusted by modifying the confining 
electric field.  

In conclusion, it was pointed out that 
two-dimensional electron systems on the surfaces of solid hydrogen or solid neon 
may have some important advantages with respect to electrons on liquid helium in 
quantum computing applications. Such quantum computers may be operational at 
higher temperatures around 4K if a sufficient smoothness of solid hydrogen or 
solid neon surfaces may be achieved. Electric field induced strong localization 
of surface electrons is shown to be a reliable tool to characterize surface 
roughness.

\end{document}